\documentclass{article}
\usepackage{spconf,amsmath,graphicx}
\usepackage{amsmath,amsfonts,amssymb,pifont}
\usepackage{adjustbox}
\usepackage{comment}
\usepackage{hyperref}
\usepackage{caption}
\usepackage{xcolor}

\usepackage{xparse}
\NewDocumentCommand\emojiclap{}{\includegraphics[scale=0.05]{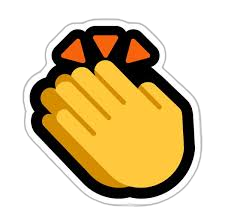}}
\usepackage{multirow,makecell}
\usepackage{setspace}

\title{NATURAL LANGUAGE SUPERVISION FOR GENERAL-PURPOSE AUDIO REPRESENTATIONS \vspace{-9pt}}

\name{Benjamin Elizalde$^{*}$, Soham Deshmukh$^{*}$, Huaming Wang   \vspace{-11pt}}
\address{\thanks{$^{*}$Equal Contribution} Microsoft \\
\{benjaminm, sdeshmukh, huawang\}@microsoft.com \vspace{-11pt}}

\begin{document}
\maketitle
\vspace{-0.05in}
\begin{abstract}
\vspace{-0.03in}

Audio-Language models jointly learn multimodal text and audio representations that enable Zero-Shot inference. Models rely on the encoders to create powerful representations of the input and generalize to multiple tasks ranging from sounds, music, and speech. Although models have achieved remarkable performance, there is still a gap with task-specific models. In this paper, we propose a Contrastive Language-Audio Pretraining model that is pretrained with a diverse collection of 4.6M audio-text pairs employing two innovative encoders for Zero-Shot inference. To learn audio representations, we trained an audio encoder on 22 audio tasks, instead of the standard training of sound event classification. To learn language representations, we trained an autoregressive decoder-only model instead of the standard encoder-only models. Then, the audio and language representations are brought into a joint multimodal space using Contrastive Learning. We used our encoders to improve the downstream performance by a large margin. We extensively evaluated the generalization of our representations on 26 downstream tasks, the largest in the literature. Our model achieves state of the art results in several tasks outperforming 4 different models and leading the way towards general-purpose audio representations. Code is on GitHub\footnote{\url{https://github.com/microsoft/CLAP}}.

\end{abstract}
\vspace{-0.1in}
\vspace{-0.05in}
\section{Introduction}\label{sec:intro}
\vspace{-0.08in}

Recent research in the audio domain focuses on learning representations that generalize to a wide range of downstream tasks across different domains. The 2021 Holistic Evaluation of Audio Representations (HEAR) \cite{turian2022hear} took a major step in this direction by providing a comprehensive setup to benchmark audio representations. The models were pretrained on a large dataset --AudioSet \cite{audioset} (1.7M files)-- using Supervised, Self-Supervised or Unsupervised Learning. All the methods have to undergo additional fine-tuning to use their representations on a given downstream task. 

Zero-Shot models can be applied to any task directly achieving flexibility and generalization. One of the most successful type are Contrastive Language-Audio Pretraining (CLAP) models that jointly learn multimodal text and audio representations. Authors in~\cite{elizalde2022clap} introduced a CLAP model that achieved state of the art (SoTA) in 16 downstream tasks. Subsequent literature showed that the choice of audio and text encoders are critical to generate powerful representations and increase performance across tasks~\cite{wu2022large, deshmukh2022audio, mei2023wavcaps, crossmodal}. For example, upgrading from CNN to audio transformers (HTSAT) to encode audio and from BERT to RoBERTa to encode text. 

Another conclusion is that scaling up the number of training pairs improves overall performance. However, simply adding pairs may result in a drop of performance in certain domains and tasks \cite{wu2022large, deshmukh2022audio, elizalde2022clap, mei2023wavcaps}. CLAP's performance is dependent on the diversity of the text and audio training pairs and how noisy they are. Wav2clip~\cite{wav2clip} and Audioclip~\cite{audioclip} used 200k and 1.7M audio-text pairs respectively from AudioSet, a dataset annotated for sound events. Authors paired audio with class labels rather than with sentence-level descriptions, potentially missing the context and language semantics of descriptions, but with good Zero-Shot performance in 3 and 9 tasks respectively. CLAP~\cite{elizalde2022clap} used 128k pairs but the text were descriptions coming from audio captioning and a web-sourced dataset. It was evaluated on 16 tasks and significantly improved over its predecessors. LAION CLAP~\cite{wu2022large} used a collection of 2.5M pairs, further improving performance in 8 tasks. Authors later added music and speech-related training pairs, but performance in sound event classification (ESC50) degraded by an absolute 1\%. Wavcaps\cite{mei2023wavcaps} used 500k pairs, but cleaned up the noisy web-sourced descriptions with a ChatGPT language model. Results outperformed the literature in 8 tasks. Therefore, when scaling up pairs it is essential to verify performance trade offs by evaluating generalization across different domains and tasks.

In this paper we make the following contributions. To learn audio representations, we trained an audio encoder on 22 audio tasks. To learn language representations, we trained an autoregressive decoder-only model. We pretrained our CLAP model with an unprecedented 4.6 million audio-text pairs and extensively evaluated the generalization of our representations on 26 downstream tasks, the largest in the literature, achieving SoTA results in several.

\vspace{-0.1in}

\vspace{-0.06in}
\section{Method}\label{sec:method}
\vspace{-0.08in}

\begin{figure}[ht]
   \centering
     \includegraphics[width=0.5\textwidth,scale=0.26, trim={0 1cm 0 4cm},clip]{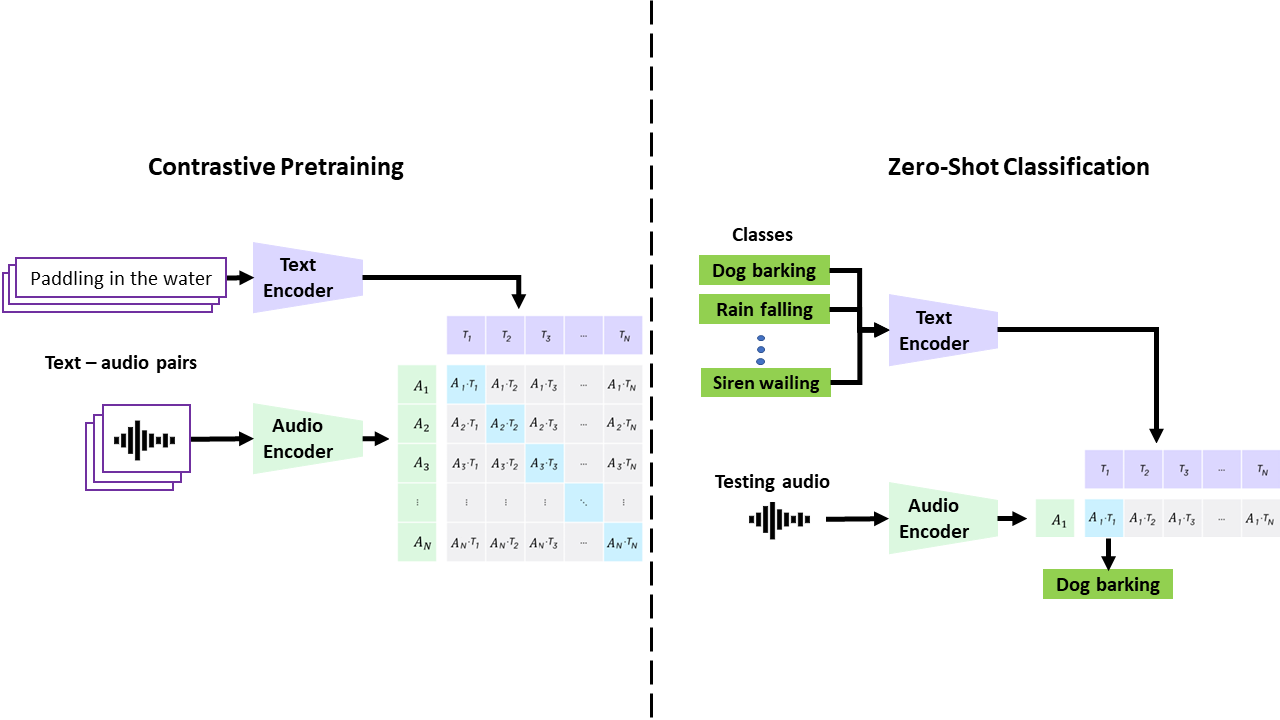}
     \includegraphics[width=0.5\textwidth,scale=0.26, trim={0 0cm 0 0cm},clip]{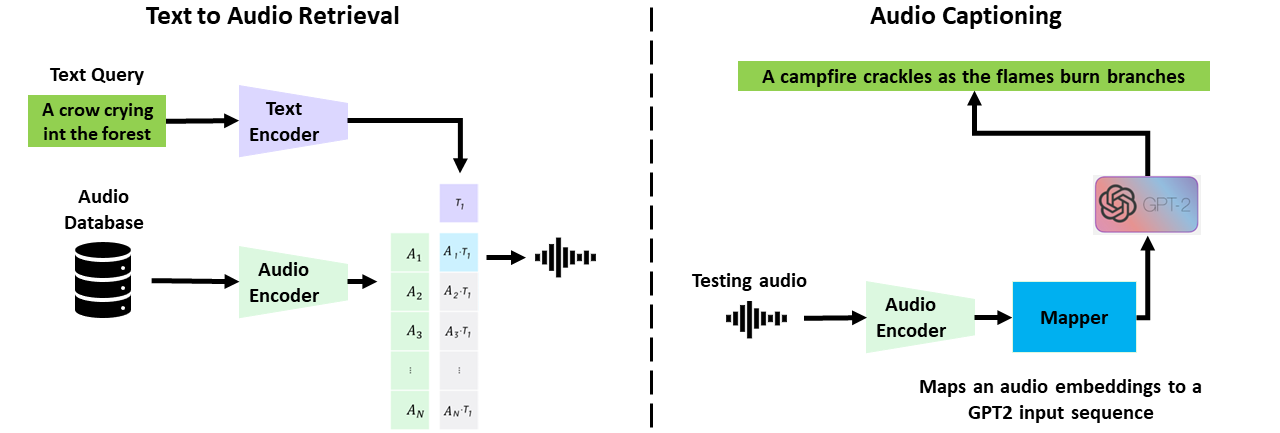}
     \caption{\small CLAP~\emojiclap{} learns audio and a text embeddings that can be compared in a multimodal space. The pretrained encoders can be used for Zero-Shot Classification, Text to Audio and Audio to Text Retrieval, and Audio Captioning. \vspace{-0.08in}}
     \label{fig:clap_diagram}
\end{figure}

Contrastive Language-Audio Pretraining (Fig~\ref{fig:clap_diagram}) jointly trains an audio an a text encoder to learn multimodal representations which can be used for different types of inference. 

\vspace{-0.1in}
\subsection{Contrastive Language-Audio Pretraining}
\vspace{-0.05in}

Let the processed audio be $X_a$ s.t. $X_a \in \mathbb{R}^{F \times T}$ where $F$ are the number of spectral components (e.g. Mel bins) and $T$ are the number of time bins. Let the text be represented by $X_t$. Each audio-text pair in a batch of $N$ is represented as $\{X_a, X_t\}_i$ where $i \in [0,N]$. For convenience, we drop the $i$ notation, and henceforth $\{X_a, X_t\}$ will denote a batch of N. 

From the pairs, the audio and text are passed to an audio encoder $f_a(.)$ and a text encoder $f_t(.)$ respectively. For a batch of N:
\begin{equation}
    \hat{X}_a = f_a(X_a); 
    \hat{X}_t = f_t(X_t)
\end{equation}
where $\hat{X}_a \in \mathbb{R}^{N \times V}$ are the audio representations of dimensionality $V$, and $\hat{X}_t \in \mathbb{R}^{N \times U}$ are the text representations of dimensionality $U$. 

We brought audio and text representations, $\hat{X}_a$ and $\hat{X}_t$, into a joint multimodal space of dimension $d$ by using a learnable projection layer: 
\vspace{-0.05in}
\begin{equation}
    E_a = L_a(X_a);
    E_t = L_t(X_t)
\end{equation}
\vspace{-0.03in}
where $E_a \in \mathbb{R}^{N \times d}$, $E_t \in \mathbb{R}^{N \times d}$, $L_a$ and $L_t$ are the projections for audio and text respectively. 

Now that the audio and text embeddings ($E_a$, $E_t$) are comparable, we can measure similarity:
\vspace{-0.05in}
\begin{equation}
    C = \tau(E_t \cdot E_a^\top)
\end{equation}
\vspace{-0.02in}
where $\tau$ is a temperature parameter to scale the range of logits. The similarity matrix $C \in \mathbb{R}^{N \times N}$ has $N$ matching pairs in the diagonal and $N^2-N$ non-matching pairs in the off-diagonal. 
\vspace{-0.05in}
\begin{equation}
     \mathcal{L} = 0.5 (\ell_{text}(C) + \ell_{audio}(C))
\end{equation}
where $\ell_{k} = \frac{1}{N}\sum_{i=0}^{N} \log diag (softmax(C))$ along text and audio axis respectively. We used this symmetric cross-entropy loss ($\mathcal{L}$) over the similarity matrix to jointly train the audio and text encoders along with their projection layers.

\vspace{-0.1in}
\subsection{Audio and Text Encoders}
\label{sec:method,subsec:encoders} 
\vspace{-0.05in}

\noindent \textbf{Audio Encoder:} To process audio, we trained a transformer-based audio encoder (HTSAT~\cite{htsat}) on 22 audio tasks using a similar method to this paper~\cite{deshmukh2023pengi}. We called it HTSAT-22. We hypothesized that an encoder trained on multiple audio tasks would improve generalization and thus performance across tasks. The method learns an audio encoder and a mapper network to prompt a large language model to perform multiple audio tasks, such as classification, captioning, retrieval and audio Q\&A. The architecture is trained essentially as a captioning system, where it learns to generate a free-form text output $c^i$ in an autoregressive fashion conditioned on the audio prompt $p^i$. Note that $\gamma$ denotes the model's trainable parameters. The loss function is Cross-Entropy:
\begin{equation}
\vspace{-0.01in}
\mathcal{L} = - \sum_{i=1}^N \sum_{j=1}^{l} \log p_{\gamma} (c^i_j| p^i_1,...,p^i_{2k}, c^i_1,...,c^i_{j-1}) 
\vspace{-0.01in}
\label{eq:captioning}
\end{equation}

\noindent \textbf{Text Encoder:} To process text, we adapted GPT2 (base 124M), which is an autoregressive model that has exhibited impressive abilities for text tasks. We addressed the challenge -- \textit{How to make an autoregressive model produce a sentence-level representation?} Autoregressive models built with transformer-decoder blocks, take an input text and output the most likely sequence of words (tokens), one after the other. In contrast, models built with transformer-encoder blocks (BERT or RoBERTA) output a sentence-level representation in a continuous space. To make GPT2 output a sentence-level representation, we appended the special token $<|endoftext|>$ at the end of each input text. During contrastive pretraining, we use the representations from this token as sentence-level representations. This forces the token to contain the aggregate information from the text input.

\vspace{-0.15in}
\subsection{Evaluation}
\label{sec:method,subsec:zeroshot} 
\vspace{-0.05in}

\noindent \textbf{Zero-Shot Inference:} We used CLAP's ability to determine the similarity between audio and text. Let's consider a target dataset with $C$ class labels and $N$ test audios. First, we compute CLAP's audio and text embeddings for $N$ audios and $C$ classes using the pretrained encoders. Second, we compute the cosine similarity between each testing audio and all the class labels. In the case of retrieval, we treat text queries as classes. Each test audio will have as many logits as classes. Third, logits are turned into a probability distribution by applying softmax for binary or multiclass classification; sigmoid for multilabel classification; and left unaltered for retrieval. 

\noindent \textbf{Audio Captioning:} In the architecture of Fig~\ref{fig:clap_diagram}, a test audio is passed to the pretrained audio encoder, then to a mapper network, and then to GPT2 to generate a description. At training time, only the weights of the mapper network are learned with a captioning loss (Eq.\ref{eq:captioning}) and the training split.

\begin{table*}[ht]
\scriptsize
\captionsetup{font=scriptsize}
\center 
\begin{tabular}{c|c|cccc|c|c|c|c} \hline
& \multicolumn{1}{c|}{\makecell{Zero-Shot\\Score $\uparrow$}} & \multicolumn{4}{c|}{Sound Event Classification $\uparrow$} & \multicolumn{1}{c|}{\makecell{Vocal Sound \\ Classification $\uparrow$}} & \multicolumn{1}{c|}{\makecell{Surveillance \\ Sound Classif.$\uparrow$}} & \multicolumn{1}{c|}{\makecell{Action\\ Classification$\uparrow$}} & \multicolumn{1}{c}{\makecell{Acoustic Scene \\ Classification$\uparrow$}}\\ \hline
\makecell{Model} & \makecell{Average} & ESC50 & FSD50K & US8K & \makecell{DCASE17 \\ Task 4} & \makecell{Vocal\\Sound} & \makecell{SESA} &\makecell{ESC50 \\ Actions} &\makecell{TUT 2017} \\ \hline 
CNN14+BERT  & 0.428 & 0.826 & 0.302 & 0.732 & 0.300 & 0.495 & 0.749 & 0.495 & 0.296\\ 
HTSAT+CLIP & 0.430 & 0.813 & 0.289 & 0.748 & 0.277 & 0.645 & 0.761 & 0.442 & 0.219\\
HTSAT+RoBERTa & 0.431 & 0.811 & 0.322 & 0.757 & 0.226 & 0.610 & 0.745 & 0.475 & 0.285\\ 
HTSAT+GPT2 & 0.435 & 0.819 & 0.336 & 0.767 & 0.242 & 0.646 & 0.644 & \textbf{0.503} & 0.286\\ \hline
HTSAT-22+RoBERTa & 0.454 & 0.879 & 0.388 & 0.767 & 0.209 & 0.682 & 0.656 & 0.481 & \textbf{0.369}\\
HTSAT-22+CLIP & 0.469 & 0.830 & \textbf{0.411} & \textbf{0.791} & 0.229 & 0.692 & 0.723 & 0.488 & 0.292\\
\textbf{HTSAT-22+GPT2} & \textbf{0.480} & \textbf{0.882} & 0.403 & 0.750 & \textbf{0.337} & \textbf{0.692} & \textbf{0.762} & 0.475 & 0.317\\ \hline
\end{tabular}
\smallskip
\vspace{-0.07in}
\center
\smallskip
\vspace{-0.07in}
\center
\begin{tabular}{c|cc|cc|cc|c|c} \hline
& \multicolumn{2}{c|}{Music Classification $\uparrow$} & \multicolumn{2}{c|}{Instrument Classification $\uparrow$} & \multicolumn{2}{c|}{\makecell{ Speech Emotion \\ Classification$\uparrow$}} & \multicolumn{1}{c|}{\makecell{KWS$\uparrow$}} & \multicolumn{1}{c}{\makecell{Speaker \\Counting$\uparrow$}} \\ \hline
\makecell{Model} & \makecell{GTZAN\\Music Speech} & \makecell{GTZAN\\Genres} & \makecell{Beijing\\Opera} & \makecell{NS Instr.\\family} & \makecell{CRE\\MA-D} & \makecell{RAV\\DESS} & \makecell{Speech \\ Commands}   & \makecell{LibriCount10} \\ \hline
CNN14+BERT & 1 & 0.252 & 0.475 & 0.295 & 0.178 & 0.160 & 0.106 & 0.179 \\ 
HTSAT+CLIP & 0.992 & 0.156 & \textbf{0.627} & 0.312 & 0.208 & 0.169 & 0.120 & 0.113 \\ 
HTSAT+RoBERTa & 0.992 & 0.178 & 0.436 & 0.352 & 0.263 & 0.2 & 0.098 & 0.149 \\ 
HTSAT+GPT2 & 1 & 0.150 & 0.539 & 0.322 & 0.234 & 0.171 & \textbf{0.139} & 0.155 \\ \hline
HTSAT-22+RoBERTa & 1 & 0.209 & 0.309 & 0.402 & 0.301 & \textbf{0.278} & 0.129 & 0.207 \\ 
HTSAT-22+CLIP & 1 & 0.280 & 0.517 & \textbf{0.462} & 0.275 & 0.233 & 0.116 & 0.094 \\ 
\textbf{HTSAT-22+GPT2} & \textbf{1} & \textbf{0.289} & 0.487 & 0.425 & \textbf{0.297} & 0.217 & 0.089 & \textbf{0.254} \\ \hline
\end{tabular}
\caption{\label{table: encoder results}\small
 Zero-Shot performance on 16 downstream tasks and 119k training pairs. Our proposed encoders (HTSAT-22+GPT2) outperformed the best combinations in the literature. Higher is better for all numbers. The metrics are mAP for FSD50k and ESC50-actions; F1-score for DCASE17; all others use Accuracy. Zero-Shot score is the average of the metrics. This is the first comparison of encoders in literature with 16 tasks, usually only a couple of enocders and a handful of tasks are considered.} \vspace{-0.2in}
\end{table*}

\vspace{-0.15in}

\section{Experiments}\label{sec:experiments}
\vspace{-0.07in}

\noindent\textbf{Training Datasets.} Collecting pairs is perhaps the main bottleneck of scaling up CLAP models. We gathered the largest collection with 4.6 million audio and text pairs from different datasets and web archives. The audios describe human sounds and activities, environmental sounds, acoustic scenes, music, sound effects, and speech emotion. To study the effect of encoders in Table ~\ref{table: encoder results}, we used the same training sets as CLAP~\cite{elizalde2022clap}. Unlike the authors, we did not include the test set of AudioCaps and Clotho, so the number of pairs was 119k instead of 128k. The training datasets for the 4.6M collection are: WavCaps~\cite{mei2023wavcaps}, AudioSet~\cite{audioset}, FSD50K~\cite{fsd50k}, Clotho~\cite{clotho}, AudioCaps~\cite{audiocaps}, MACS~\cite{macs}, WavText5k~\cite{deshmukh2022audio}, SoundDesc~\cite{sounddescs}, NSynth~\cite{nsynth}, FMA~\cite{fma}, Mosi~\cite{mosi}, Meld~\cite{meld}, Iemocap~\cite{iemocap}, Mosei~\cite{mosei}, MSP-Podcast~\cite{msp_podcast}, CochlScene~\cite{jeong2022cochlscene}, LJspeech~\cite{ljspeech17}, EpicKitchen~\cite{epickitchen}, Kinectics700~\cite{kinetics700}, findsounds.com. Details on GitHub. 

\noindent\textbf{Downstream Tasks.} We used 26 downstream tasks from different domains, several come from HEAR\cite{turian2022hear}: sound events, vocal sounds, surveillance sounds, and acoustic scenes classification; audio captioning; retrieval; music, instruments, and note attributes classification; speech emotions and language classification; keyword spotting; and speaker counting. To study the effect of encoders in Table ~\ref{table: encoder results}, we used a subset of 16 tasks. 

\noindent\textbf{Pre-processing.} We used log Mel spectrogram representations of audio with a sampling rate of 44.1 KHz, hop size of 320 frames, window size 1024 frames, and 64 Mel bins in the range of 50-8000 Hz. During training, each audio clip is randomly truncated to a continuous segment of 7 secs, or padded if shorter. The batches with pairs are randomly sampled. 

\noindent\textbf{Encoders.} For our proposed CLAP model, we used the audio and text encoders HTSAT-22+GPT2 described in Sec.\ref{sec:method,subsec:encoders}. For comparison, in Table~\ref{table: encoder results} we used the two best combinations of encoders in the literature CNN14+BERT and HTSAT+RoBERTa~\cite{elizalde2022clap,wu2022large, mei2023wavcaps}. We also included the text encoder from CLIP because it was used by different authors~\cite{audioclip,wav2clip,wu2022large}. Both, the audio and text embeddings are projected into a multimodal space with independent learnable projection layers with an output dimension of 1024. 

\noindent\textbf{Training.} We trained by unfreezing both encoders for 40 epochs, although the overall performance peaked in the first 10 epochs. We report the performance of the downstream tasks corresponding to the epoch that yielded the best Zero-Shot score (average of all tasks). We hypothesize that the model corresponding to such epoch will generalize better to unseen datasets and serve the community better. It is possible that the performance of each task was higher or lower in a different epoch. Batch size was 1,536. We used Adam Optimiser with an initial learning rate $10^{-3}$ and reduce the learning rate on plateau by $10^{-1}$ with a patience of 15. The temperature parameter $\tau$ is learnable and initialised to 0.007. 

\vspace{-0.1in}

\vspace{-0.07in}
\section{Results and Discussion}\label{sec:results}
\vspace{-0.1in}

\begin{table*}[ht]
\scriptsize
\captionsetup{font=scriptsize}
\center 
\begin{tabular}{c|ccccc|c|c|c|c} \hline 
& \multicolumn{5}{c|}{Sound Event Classification $\uparrow$} & \multicolumn{1}{c|}{\makecell{Vocal Sound \\ Classification $\uparrow$}} & \multicolumn{1}{c|}{\makecell{Surveillance \\ Sound Classif.$\uparrow$}} & \multicolumn{1}{c|}{\makecell{Action\\ Classification$\uparrow$}} & \multicolumn{1}{c}{\makecell{Acoustic Scene \\ Classification$\uparrow$}}\\ \hline
\makecell{Model} & \makecell{ESC50 \\ ~\cite{esc50}}\ & \makecell{FSD50K \\ ~\cite{fsd50k}} & \makecell{US8K \\~\cite{UrbanSound}} & \makecell{DCASE17 \\ Task 4~\cite{mesaros2019sound}} & \makecell{AudioSet \\~\cite{audioset}} & \makecell{Vocal\\Sound~\cite{vocalsound}} & \makecell{SESA \\~\cite{sesa}} &\makecell{ESC50 \\ Actions~\cite{esc50actions}} &\makecell{TUT 2017 \\~\cite{mesaros2019sound}} \\ \hline
Benchmark &  \textbf{0.948}~\cite{mei2023wavcaps} & 0.302~\cite{elizalde2022clap} & 0.806~\cite{mei2023wavcaps} & 0.3~\cite{elizalde2022clap} & 0.058~\cite{elizalde2022clap} & 0.495~\cite{elizalde2022clap} & 0.25 & 0.045 & 0.296~\cite{elizalde2022clap}\\ 
\textbf{HTSAT-22+GPT2} & 0.939 & \textbf{0.485} & \textbf{0.823} & \textbf{0.466} & \textbf{0.102} & \textbf{0.8} & \textbf{0.65} & \textbf{0.509} & \textbf{0.538}\\ \hline
\end{tabular}
\smallskip
\vspace{-0.05in}
\center
\smallskip
\vspace{-0.07in}
\center
\begin{tabular}{c|ccccc|cc|cc|c|c} \hline
& \multicolumn{5}{c|}{Music Classification $\uparrow$} & \multicolumn{2}{c|}{Instrument Classification $\uparrow$} & \multicolumn{2}{c|}{\makecell{ Speech Emotion \\ Classification$\uparrow$}} & \multicolumn{1}{c|}{\makecell{KWS$\uparrow$}} & \multicolumn{1}{c}{\makecell{Speaker \\Counting$\uparrow$}} \\ \hline
\makecell{Model} & \makecell{GTZAN\\Music \\Speech~\cite{turian2022hear}} & \makecell{GTZAN\\Genres\\~\cite{turian2022hear}} & \makecell{NS \\ Pitch\\~\cite{nsynth}} & \makecell{NS \\ Velocity\\~\cite{nsynth}} & \makecell{NS \\ Qualities\\~\cite{nsynth}} & \makecell{Beijing\\Opera\\~\cite{turian2022hear}} & \makecell{NS Instr.\\family\\~\cite{nsynth}}  & \makecell{CRE\\MA-D\\~\cite{turian2022hear}} & \makecell{RAV\\DESS\\~\cite{ravdess}} & \makecell{Speech \\ Commands \\~\cite{turian2022hear}}   & \makecell{Libri \\ Count10\\~\cite{turian2022hear}} \\ \hline
Benchmark & \textbf{1}~\cite{elizalde2022clap} & 0.25~\cite{elizalde2022clap} & 0.015 & 0.2 & 0.1 & \textbf{0.4746}~\cite{elizalde2022clap} & 0.09 & 0.178~\cite{elizalde2022clap} & 0.159~\cite{elizalde2022clap} & 0.106~\cite{elizalde2022clap} & 0.178~\cite{elizalde2022clap} \\ 
\textbf{HTSAT-22+GPT2} & 0.992 & \textbf{0.584} & \textbf{0.444} & \textbf{0.222} & \textbf{0.489} & 0.466 & \textbf{0.479} & \textbf{0.3} & \textbf{0.315} &\textbf{0.164} & \textbf{0.246}\\ \hline
\end{tabular}
\smallskip
\vspace{-0.05in}
\center
\smallskip
\vspace{-0.07in}
\center
\begin{tabular}{c|cc|cccc|cccc} \hline
& \multicolumn{2}{c|}{\makecell{Audio Captioning $\uparrow$}} & \multicolumn{4}{c|}{\makecell{Audio-Text Retrieval $\uparrow$}} & \multicolumn{4}{c}{\makecell{Text-Audio  Retrieval $\uparrow$}}  \\ \hline 
\makecell{Model} & \makecell{AudioCaps\\~\cite{audiocaps}} & \makecell{Clotho\\~\cite{clotho}} & \makecell{AudioCaps \\ R@1} & \makecell{AudioCaps \\ mAP@10} & \makecell{Clotho \\ R@1} & \makecell{Clotho \\ mAP@10} & \makecell{AudioCaps \\ R@1} & \makecell{AudioCaps \\ mAP@10} & \makecell{Clotho \\ R@1}& \makecell{Clotho \\ mAP@10} \\ \hline
Benchmark & 0.438\cite{kim2023prefix} & 0.215\cite{kim2023prefix} & \textbf{0.517}\cite{mei2023wavcaps} &\textbf{0.457 }\cite{wu2022large} & \textbf{0.234}\cite{mei2023wavcaps} & 0.138\cite{wu2022large}  & \textbf{0.397}\cite{mei2023wavcaps} & 0.51\cite{wu2022large} & \textbf{0.195}\cite{mei2023wavcaps} & 0.204\cite{wu2022large}\\ 
\textbf{HTSAT-22+GPT2} & \textbf{0.455} & \textbf{0.271} & 0.425 & 0.319   &  0.229 & \textbf{0.155} & 0.356  & \textbf{0.51} & 0.157 & \textbf{0.257}\\ 
\hline
\end{tabular}
\caption{\label{table: Zero-shot results}\small Performance on 26 downstream tasks using our proposed encoders and 4.6M training pairs. As the benchmark, we used the best numbers in the literature, when no number was available we used random performance. Higher is better for all tasks. The evaluation metrics are mAP for FSD50k, ESC50-Actions, AudioSet, and NS Qualities; F1-score for DCASE17; and SPIDEr for Captioning; all others use Accuracy.} \vspace{-0.2in}
\end{table*}

The results comparing different audio and text encoders are in Table~\ref{table: encoder results} and the results of our proposed CLAP are in Table~\ref{table: Zero-shot results}. 

\vspace{-0.15in}
\subsection{Proposed audio and text encoder}\label{sec,results,subsec:encoders}
\vspace{-0.05in}

Our proposed encoders HTSAT-22+GPT2 outperformed two of the best combination of encoders in the literature, as shown in Table~\ref{table: encoder results}. To compare overall performance, we used Zero-Shot score, which is the average of the metrics from all 16 tasks. HTSAT-22+GPT2 achieved 0.480, an absolute 9\% higher than the most common combinations HTSAT+RoBERTa and CNN14+BERT with 0.431 and 0.428 respectively. All encoder combinations performed better than random. Although different combinations did better at different tasks, none of them excelled at a specific domain.

Our HTSAT-22 audio encoder is the major contributor to performance improvement. HTSAT-22 is pretrained on 22 audio tasks in contrast to HTSAT which is pretrained only on sound event classification. Hence, suggesting that generating pretraining on multiple audio tasks can improve the representations from the audio encoder. Comparing HTSAT-22+GPT2 to HTSAT+GPT2 evidenced major improvements such as LibriCount10 (absolute 10\%), NS Instrument (absolute 7\%) and ESC50 (absolute 6\%). 

The proposed GPT2 autoregressive model improves upon the popular RoBERTa. Using GPT2 with either HTSAT or HTSAT-22 yielded the best performance over the other text encoders. We hypothesize that the improvement comes from two reasons. First, GPT2 has a larger vocabulary of 50k tokens compared to BERT and RoBERTa with 30k. Second, our modified GPT2 autoregressive predicts tokens till $<|endoftext|>$ used for sentence-level representation. This acts as self-supervision and forces the model to learn and put emphasis on the ordering of words.

\vspace{-0.15in}
\subsection{Scaling proposed CLAP architecture}\label{sec:results,subsec:zero shot results}
\vspace{-0.05in} 

Our CLAP model established new Zero-Shot SoTA on most of the 26 downstream tasks as shown in Table~\ref{table: Zero-shot results}, outperforming 4 different SoTA models. To benchmark our model, we used the best numbers in the literature coming from different models. When no number was available, we used random performance. In some cases, performance improvement is more than double the benchmark literature. Some highlights are Music Genres with 58.4\% acc. vs 25\%, Vocal Sounds with 80\% acc. vs 49.5\%, Acoustic Scenes with 53.8\% acc. vs 29.6\%. Some downstream tasks do not constitute a true Zero-Shot setup as the audio files in the training set were part of the 4.6M pairs (see Sec.\ref{sec:experiments}). For instance, FSD50k audio and web descriptions were used in training but not the class labels. We did not fine-tune CLAP encoders for any downstream task. We only fine-tune the audio encoder for ESC50 and were able to improve performance from our previous CLAP model from 96.70\% to 98.25\% accuracy, thus establishing a new SoTA.

\vspace{-0.15in}
\subsection{Generalization and individual domain performance}\label{sec:results,subsec:tradeoff}
\vspace{-0.05in} 

Adding diversity and scaling the audio-text pairs in training presents a trade-off that increases performance in some tasks but decreases it in others. As expected, adding training pairs that resemble the domain from a given task helps, hence diversity is essential for generalization. For example, CLAP~\cite{elizalde2022clap} did not include emotion recognition training pairs and achieved 17.1\% acc. in RAVDESS and 23.4\% in CREMAD. We added emotion-related pairs and improved accuracy to 31.5\% and 30\% respectively. Nonetheless, more pairs can cause a distribution shift, creating a mismatch between training and some testing data. For example, our model achieved a slightly lower score than a model~\cite{mei2023wavcaps} trained with 500k pairs on ESC50 (94.8\% vs 93.9\% acc.). Another example is with GTZAN Music vs Speech, where a model~\cite{elizalde2022clap} with 128k pairs achieved 100\% acc. over ours with 99.2\%. Even our model in Table~\ref{table: encoder results} achieved 100\% acc with 119k pairs. We should expect that as we add training pairs, performance across tasks will vary. Hence, zero-shot models should be evaluated across different domains and tasks with focus on generalization rather than on overfitting to specific tasks. 

Audio-Text (A-T) and Text-Audio (T-A) Retrieval performance fell short of the benchmark. We measured the tasks with mAP@10, which is the ranking metric of IEEE DCASE, and R@1. Our model outperformed the literature in terms of mAP@10 for Clotho (A-T: 0.155 vs 0.138 and T-A: 0.257 vs 0.204), and struggled only with A-T AudioCaps (A-T: 0.319 vs 0.457 and T-A: 0.51 vs 0.51). Both datasets are sensitive to out-of-domain training data and adding training pairs did not translate into an improvement. This was demonstrated by authors in~\cite{deshmukh2022audio} who unsuccessfully tried to add 39k files from SounDesc or authors in~\cite{wu2022large} with 500k from Wavcaps or authors in~\cite{mei2023wavcaps} with 1.7M from AudioSet.  
\vspace{-0.2in}

\vspace{-0.03in}
\section{Conclusion}\label{sec:conclusions}
\vspace{-0.07in}
We introduced a CLAP model with our proposed encoders and 4.6M training pairs. Zero-shot models should be evaluated across different tasks with a focus on generalization rather than on overfitting to specific tasks. We evaluated CLAP on 26 tasks and established SoTA on most of them, leading the way in general-purpose audio representations.


\begin{spacing}{0.8}\bibliographystyle{IEEEbib}
\bibliography{refs}

\end{spacing}
\end{document}